\pdfoutput=1 

\documentclass[prd,aps,twocolumn,amsmath,nofootinbib,amssymb,eqsecnum,showkeys,tightenlines]{revtex4-1}

\usepackage{amsmath, latexsym, amssymb, hyperref, graphicx, color, multirow}
\usepackage{dsfont}
\usepackage[table]{xcolor}
\usepackage{tabularx}
\usepackage[T1]{fontenc}
\usepackage[capitalize]{cleveref}
\definecolor{nicered}{rgb}{.7,.1,.1}
\definecolor{nicegreen}{rgb}{.1,.5,.1}
\definecolor{darkblue}{rgb}{0,0,.5}
\hypersetup{colorlinks, citecolor=nicegreen,linkcolor=nicered, urlcolor=darkblue}
\usepackage{multirow}
\usepackage{slashed}
\numberwithin{equation}{section}
\usepackage{verbatim}

\begin{document}

\title{Uncovering quirk signal via energy loss inside tracker}

\author{Jinmian Li$^{1}$}
\email{jmli@scu.edu.cn}

\author{Tianjun Li$^{2,3}$}
\email{tli@mail.itp.ac.cn}

\author{Junle Pei$^{2,3}$}
\email{peijunle@mail.itp.ac.cn}

\author{Wenxing Zhang$^{2,3}$}
\email{zhangwenxing@mail.itp.ac.cn}

\affiliation{$^1$ College of Physics, Sichuan University, Chengdu 610065, China}
\affiliation{$^2$CAS Key Laboratory of Theoretical Physics, Institute of Theoretical Physics, Chinese Academy of Sciences, Beijing 100190, China}
\affiliation{$^3$ School of Physical Sciences, University of Chinese Academy of Sciences,
No.~19A Yuquan Road, Beijing 100049, China}

\begin{abstract}
A quirk propagating through a detector is subject to the Lorentz force, a new confining gauge force, and the frictional force from ionization energy loss. At the LHC, it was found that the monojet search and the coplanar search were able to constrain such a quirk signal. 
Inspired by the coplanar search proposed by S. Knapen et.al.	, we develop a new search that also utilizes the information of the relatively large ionization energy loss inside tracker. 
Our algorithm has improved efficiency in finding quirk signals with a wide oscillation amplitude. 
Because of our trigger strategy, the $Z(\to \nu\nu)+$jets process overlaid by pileup events is the dominant background. We find that the $\sim 100$ fb$^{-1}$ dataset at the LHC will be able to probe the colored fermion(scalar) quirks with masses up to {2.1(1.1) TeV}, and the color neutral fermion(scalar) quirks with masses up to {450(150) GeV}, respectively. 

 \end{abstract}

\maketitle

\section{Introduction}

The Large Hadron Collider (LHC) has already collected tremendous data at its Run-1 and Run-2 
with the center of mass energy ranging from 7 TeV to 13 TeV. However, the traditional analyses failed 
to find any new physics signals beyond the standard model (BSM) from those data. 
Yet there still exists the possibility that some new physics processes have been copiously produced 
at the LHC without being probed because of their non-conventional behavior. The long-lived exotic searches 
are receiving increased interests at the LHC~\cite{Lee:2018pag,Alimena:2019zri} and 
some future facilities~\cite{Chou:2016lxi, Ariga:2019ufm}. 
Many BSM scenarios, which include extra gauge symmetries (such as hidden valley models~\cite{Strassler:2006im,Cheng:2019yai}), predict exotic signals at the detector, for example, emerging jet~\cite{Schwaller:2015gea}, trackless jet~\cite{Daci:2015hca}, and soft bomb~\cite{Knapen:2016hky}, etc. 
Even though those exotic signals were overlooked by traditional searches, it does not mean that they are difficult to probe. In fact, there are already specific searches designed for emerging jet~\cite{Sirunyan:2018njd} and disappearing track~\cite{Sirunyan:2018ldc} by the CMS Collaboration. 
Very stringent bounds on those signals were obtained because of the low backgrounds.

Quirks are long-lived exotic particles that are charged under both the Standard Model (SM) gauge group 
and a new confining gauge group. The mass of the lightest quirk is much larger than the confinement scale ($\Lambda$) of the new gauge group. 
In this paper, we shall consider the quirk pair production signals at the LHC~\cite{Okun:1980mu,Kang:2008ea}. 
Due to the confining gauge force, two quirks will start to oscillate after production. 
The typical oscillation amplitude in the center of mass frame can be estimated as 
\begin{align}
\ell = 1_{\text{cm}} \times \Theta (\frac{1~\text{keV}}{\Lambda})^2 \times  \Theta(\frac{m_{\mathcal{Q}}}{100~\text{GeV}})~, \label{eq:amp2}
\end{align}
where $m_{\mathcal{Q}}$ is quirk mass. 
Giving quirk mass around the electroweak scale, when  $\Lambda \gtrsim \mathcal{O}(10)$ MeV, the confining gauge force will lead to intensive quirk oscillation. The quirk pair will lose energy quickly via photon and hidden glueball radiation. As a result, the quirk pair annihilates into the SM particles almost promptly($\ll 1$ ns). 
Searches for new resonances in the SM final states have been proposed to probe such quirk signals~\cite{Cheung:2008ke,Harnik:2011mv,Fok:2011yc,Curtin:2015jcv,Chacko:2015fbc}. 
For $\Lambda \in [10~\text{keV}, 10~\text{MeV}]$, the quirk oscillation amplitude is microscopic (undetected by the detector). Meanwhile, the glueball and photon radiations are not frequent enough so that the quirk pair can be long lived~\cite{Kang:2008ea}. 
Since the quirk pair system is electric neutral, the quirk hits on the tracker almost lie on a straight line.  Those hits can be reconstructed as a single boosted charged particle with high ionization energy loss. This signal has been searched at Tevatron~\cite{Abazov:2010yb}. 
As for very small $\Lambda \lesssim \mathcal{O}(10)$ eV, the confining force becomes negligible comparing to the Lorentz force. 
In this case, the trajectory of each quirk is still a helix.  Such signals will be constrained by heavy stable charged particle searches at the LHC~\cite{CMS-PAS-EXO-16-036,Aaboud:2016uth,Farina:2017cts}.

This work will focus on the case with $\Lambda \sim [100~\text{eV}$-$10~\text{keV}]$, where the oscillation amplitude is macroscopic and each quirk trajectory can not be simply reconstructed as a helix. Moreover, the quirk pair can deposit only a little energy in the electromagnetic/hadronic calorimeter (ECal/HCal) within the timescale of 25 ns (which is  the bunch crossing period of the LHC). 
Both quirks will be missed by conventional reconstructions in collider searches and will just behave as missing transverse energy. As a result, the quirk signal can be constrained only by monojet searches~\cite{CMS-PAS-EXO-16-037,Aaboud:2016tnv,Farina:2017cts} when they are boosted by recoiling against an energetic initial state radiated jet. 
On the other hand, if the quirk pair is produced with little kinetic energy, the ECal or HCal is able to capture the quirk system. The quirk pair will eventually annihilate inside the calorimeter at some time when there are no active $\mathds {pp}$ collisions.~\cite{Evans:2018jmd}. 

However, the quirk signal ($\Lambda \sim 1~\text{keV}$, $m\sim \mathcal{O}(100)$ GeV) could be more informative than just missing transverse energy. For a boosted quirk pair, the dominant confining force will lead to coplanar trajectories. Searching for coplanar hits in the tracker can greatly suppress the backgrounds while maintain very high signal efficiency. 
We further develop the coplanar quirk search as proposed in Ref.~\cite{Knapen:2017kly}  by adding the information of ionization energy loss ($dE/dx$) inside the tracker. 
The $dE/dx$ of each cluster generated by the charged particles throughout the detector can be derived from the cluster charge, the average energy in creating an electron-hole pair, the density of silicon, and the thickness of each layer~\cite{ATLAS:2011gea}. 
Our method relies on the fact that quirks leave hits with relatively larger $dE/dx$ in the tracker than the SM particles~\cite{Khachatryan:2016sfv}.
Aiming to search quirk signals with oscillation amplitude less than $\mathcal{O}(1)$ cm, the parameters in the analysis of Ref.~\cite{Knapen:2017kly} are chosen such that the time complexity of their algorithm is $\mathcal{O}(10^6)$. However, this number grows as $\ell^4$, with $\ell$ given in Eq.(\ref{eq:amp2}). 
 Using the information of $dE/dx$, the coplanar search algorithm can be much more efficient, especially when the oscillation amplitude exceeds $\mathcal{O}(1)$ cm. 
We will show later that the time complexity of our algorithm is $\mathcal{O}(10^6)$, which, however, is insensitive to the quirk oscillation amplitude.
Moreover, we find that the $Z(\to \nu\nu)+$jets process overlaid by pileup events is the dominant background since the signal is triggered by large missing transverse energy~\footnote{The $Z(\to \nu\nu)+$jets background is not considered in Ref.~\cite{Knapen:2017kly}, which will lead to overoptimistic results. We will give a more detailed discussion in the conclusion.}.  

This paper is organized as follows. In Sec.~\ref{sec:th}, we briefly discuss the theoretical frameworks  for the quirk.  Detailed studies on the equation of motion for quirks will be given in Sec.~\ref{sec:motion}. In Sec.~\ref{sec:ana}, based on Monte Carlo events, we propose the method to separate the quirk signal from backgrounds. We draw the conclusion in Sec.~\ref{sec:conc}.

\section{The nature of quirks}
\label{sec:th}
The color neutral quirks are commonly present in many BSM models of neutral naturalness~\cite{Curtin:2015bka}, which are proposed to partly solve the little hierarchy problem~\cite{BasteroGil:2000bw,Bazzocchi:2012de}. Such models 
include folded supersymmetry~\cite{Burdman:2006tz,Burdman:2008ek}, quirky little Higgs~\cite{Cai:2008au}, twin Higgs~\cite{Chacko:2005pe,Craig:2015pha,Serra:2019omd}, minimal neutral naturalness model~\cite{Xu:2018ofw}, and so on, while other more general BSMs predict quirks that carry strong interaction~\cite{Martin:2010kk}. The quirk can be either a fermion or a scalar as well.

In this work, we will take the simplified model frameworks as benchmark. The quantum numbers of the quirks under $SU(N_{\text{IC}}) \times SU_C(3) \times SU_L(2) \times U_Y(1)$ are given as 
\begin{align}
 \tilde{\mathcal{D}}^c &= \left( N_{\text{IC}}, 3, 1, 2/3 \right)~,~~\\
 \tilde{\mathcal{E}}^c &= \left( N_{\text{IC}}, 1, 1, -2 \right)~,~~\\
 \mathcal{D}^c &= \left( N_{\text{IC}}, 3, 1, 2/3 \right)~,~~\\
\mathcal{E}^c &= \left( N_{\text{IC}}, 1, 1, -2 \right) ~,
\end{align}
where we take $N_{\text{IC}}=3$ for the infracolor gauge group.  $\tilde{\mathcal{D}}^c$ and $\tilde{\mathcal{E}}^c$ are spin zero particles, while $\mathcal{D}^c$ and $\mathcal{E}^c$ are fermions. 
The electric charges of  $\tilde{\mathcal{D}}^c$/$\mathcal{D}^c$ and $\tilde{\mathcal{E}}^c$/$\mathcal{E}^c$ are $\frac{1}{3}$ and -1, respectively. However, due to the color confinement,  one can observe only the quirk-quark bound state for  $\tilde{\mathcal{D}}^c$ and $\mathcal{D}^c$.  The probability for the quirk-quark bound state to have charge $\pm 1$ is around 30\%~\cite{Knapen:2017kly}. 
Since we are interested only in quirk-quark bound states with nonzero electric charges, in the following, we will simply refer to the charge $\pm 1$ quirk-quark bound state as  $\tilde{\mathcal{D}}^c$ or $\mathcal{D}^c$. 

\begin{figure}[thb]
\begin{center}
\includegraphics[width=0.5\textwidth]{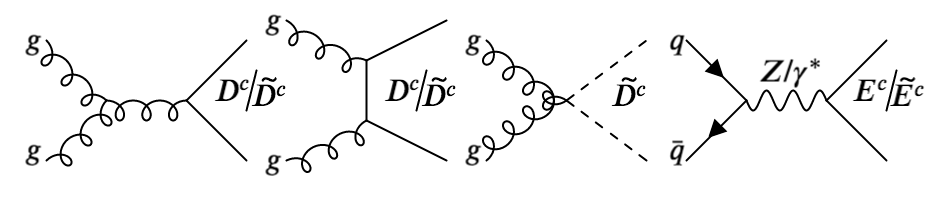}
\end{center}
\caption{\label{fig:proc} Production processes of quirks with different representations at the LHC. }
\end{figure}

The quirks can be pair produced at colliders through SM gauge interaction. 
The dominant production channels for different quirks are shown in Fig~\ref{fig:proc}.
The colored fermionic quirk production is given by the first two diagrams, while the colored scalar quirk receives an extra contribution from the third diagram. The color neutral  quirk production is simply given by the Drell-Yan processes in the fourth diagram. 

\begin{figure}[thb]
\begin{center}
\includegraphics[width=0.5\textwidth]{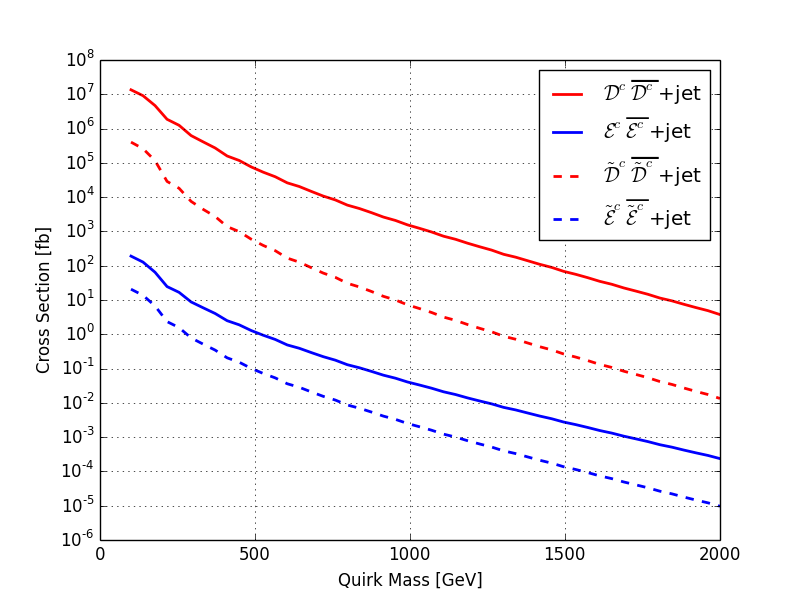}
\end{center}
\caption{\label{fig:xsec} The leading-order production cross sections for different quirks at the 13 TeV LHC. We have also required an ISR  jet with $p_t(\text{jet})>100$ GeV in production processes. }
\end{figure}

We also provide the production cross sections for these quirks at the 13 TeV LHC in Fig.~\ref{fig:xsec}. 
The quirk pair events with coplanar trajectories can be triggered by large missing transverse energy ($E^{\text{miss}}_T \gtrsim 100$ GeV~\cite{CMS-PAS-EXO-16-037}), which is induced by an energetic initial state radiation (ISR) in quirk production. So we have imposed the cut $p_T>100$ GeV on the ISR jet in calculating the cross sections. 
Because the scalar quirks have fewer degrees of freedom than the fermionic quirks and their couplings to gauge bosons are momentum suppressed, the cross sections of scalar quirks are typically more than one order of magnitude smaller than those of fermionic quirks, given the same mass and gauge group representation.

\section{Quirk equation of motion}
\label{sec:motion}
The quirk equation of motion (EOM) inside the detector is given by 
\begin{align}
\frac{\partial (m \gamma \vec{v})}{\partial t} &=\vec{F}_{s}+\vec{F}_{ext}~,\label{eq::move}\\
\vec{F}_{s}&=-\Lambda^2\sqrt{1-\vec{v}_{\perp}^{2}} \hat{s}-\Lambda^2 \frac{v_{ \|} \vec{v}_{\perp}}{\sqrt{1-\vec{v}_{\perp}^{2}}}~,\label{eq::fs} \\ 
\vec{F}_{ext} &=  q \vec{v} \times \vec{B} - \langle \frac{dE}{dx} \rangle \hat{v}~,
\end{align}
where $v_{ \|}=\vec{v}\cdot\hat{s}\nonumber$ and $\vec{v}_{\perp}=\vec{v}-v_{ \|}\hat{s}\nonumber$ with 
$\hat{s}$ being a unit vector that points toward the other quirk in the center of mass frame. 
$\vec{F}_{s}$ corresponds to the infracolor force and is determined by the Nambu-Goto action~\cite{Luscher:2002qv}. 
$\vec{F}_{\text{ext}}$  represents the external forces, which includes the Lorentz force and the frictional force from ionization energy loss. 

There are several subdominant effects that are not taken into account in the EOM. A colored quirk is surrounded by a cloud of nonperturbative QCD ``brown muck''~\cite{Kang:2008ea}. Because of the nonperturbative QCD interaction, every time two quirks cross each other during the oscillation, a hadron with energy $\sim \Lambda_{\text{QCD}}$ will be radiated. Similarly, two quirks bound by infracolor string can emit soft infracolor glueballs with energy roughly of the order of $\Lambda$. 
At last, the energy loss due to Larmor radiation is proportional to $\sim \sqrt{\alpha} \Lambda \sim 0.1 \Lambda$.

\subsection{The $dE/dx$ inside the CMS detector}
\label{sec:dedx}

As a function of velocity in the Bethe-Bloch (BB, $\beta \gamma \gtrsim 0.06$) and Lindhard-Scharff (LS, $\beta \gamma \lesssim 0.004$) regions, the average ionization energy loss of charged particles is well 
predicted by the following equations~\cite{Chilingarov:1999vi,Schou:2006zz}: 
\begin{align}
\langle \frac{dE}{dx}(v)_{\text{LS}}  \rangle &=  A_1 v, \label{dedx1}\\
\langle \frac{dE}{dx}(v)_{\text{BB}} \rangle&= A_2 \frac{q^2}{v^2} \ln \left( \frac{A_3 v^2}{1-v^2} -v^2 \right), \label{dedx2}
\end{align}
where the coefficients are given by
\begin{align} 
A_1 &=(3.1 \times 10^{-11}~\text{GeV}^2)  \frac{\rho}{g/\text{cm}^3}  \frac{q^{7/6} Z}{A (q^{2/3}+Z^{2/3})^{3/2}},  \nonumber\\
A_2 &= (6.03 \times 10^{-18}~\text{GeV}^2)  \frac{\rho}{g/\text{cm}^3}  \frac{Z}{A}, \nonumber \\
A_3 &= \frac{102200}{Z}, \nonumber
\end{align}
with the $A$, $Z$ and $\rho$ corresponding to the relative atomic mass, atomic number and density of material, respectively.
The ionization energy loss function in the region between LS and BB is interpolated from experimental data. 
In a realistic measurement, the fluctuations of the ionization energy loss will follow Landau-Vavilov distribution in thick sensors~\cite{Landau:1944if,Vavilov:1957zz} and Bichsel distribution in thin sensors~\cite{Bichsel:1988if,Wang:2017ygj}. The thickness of tracking layers in the CMS detector ranges from 280 to 500 $\mu$m~\cite{Khachatryan:2010pw}, so the energy fluctuation can be well described by the Landau-Vavilov distribution.
In our study, we take a Gaussian smearing on the $\langle dE/dx \rangle$ with the uncertainty about 10\% for simplicity~\cite{Khachatryan:2010pw}. 

\begin{figure}[thb]
\begin{center}
\includegraphics[width=0.5\textwidth]{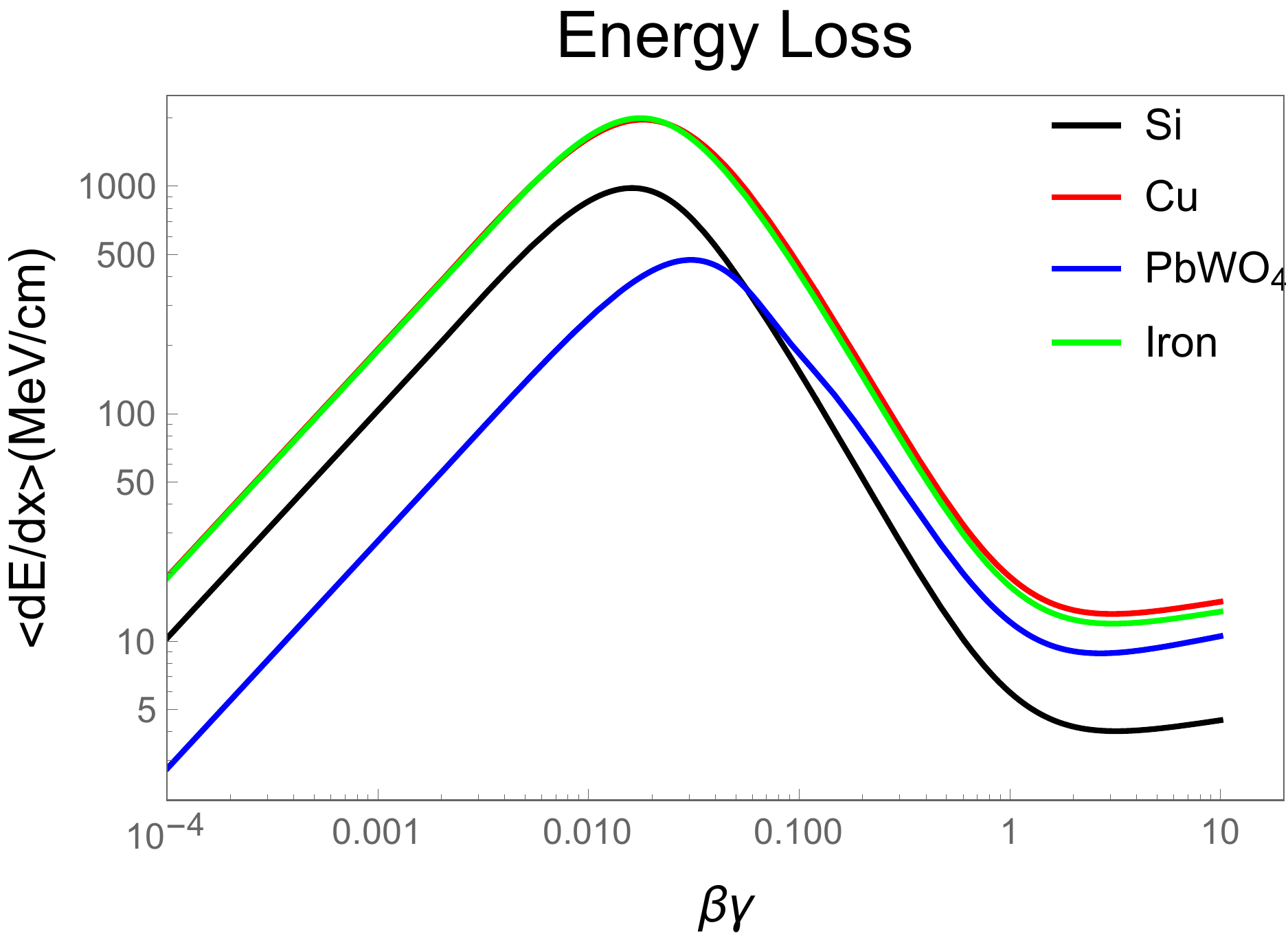}
\end{center}
\caption{\label{fig:dedx} The ionization energy loss for charge $q=1$ particle as a function of $\beta \gamma = v/\sqrt{1-v^2}$ in different materials. }
\end{figure}

The CMS detector uses silicon material for the tracker layers, lead tungstate (PbWO$_4$) for the ECal, copper for the HCal, and iron for the muon chamber. Taking parameters of materials from the Particle Data Group~\cite{Tanabashi:2018oca}, we plot the ionization energy loss function for the quirk propagating through each of those materials in Fig.~\ref{fig:dedx}. 
It has to be noted that the detector volume is not fully occupied. The filling rates for the tracker, ECal and HCal of the CMS detector are 0.05, 0.33 and 0.88, respectively. The ionization energy loss functions in the figure should be rescaled by the corresponding factors in solving the quirk EOM inside the CMS detector. 

\subsection{Numerical solution of the EOM}

The tracker records the positions of quirk hits in the laboratory (lab) frame. However, the dynamics of quirk pair in the lab frame is much more difficult to describe than that in the center of mass (c.m.) frame. 
In the c.m. frame, the infracolor string is straight. So the direction of the $\hat{s}$ for one quirk is simply given by the relative displacement respect to the other quirk. 
However, the c.m. frame itself is changing all the time due to the existence of the $\vec{F}_{\text{ext}}$.  
Since the $\vec{F}_{\text{ext}}$ depends on the velocity of each quirk, the simultaneity and
collineation of the infracolor force between two quirks in the lab frame no longer persist. 

We numerically solve the quirk EOM by Euler's method with a small time step. In the lab frame,  the four-momenta of the two quirks are denoted by $(E_i, \vec{P}_i)$, and the space-time positions are denoted by  $(t_i, \vec{r}_i)$ with $i=1,2$. In order to ensure the simultaneity in the c.m. frame, the following condition is required
\begin{equation}\label{eq::synchrone}
t_1-t_2=\vec{\beta}\cdot(\vec{r}_1-\vec{r}_2)
\end{equation}
with $\vec{\beta}=(\vec{P}_1+\vec{P}_2)/(E_1+E_2)$.
As a result, the time step sizes for two quirks $\epsilon_i (\ll 1)$  should satisfy
\begin{align}\label{eq::length}
\epsilon_{1}[1- & \vec{v}_{1} \cdot \vec{\beta}-\frac{\vec{r}_{1}-\vec{r}_{2}}{E_{1}+E_{2}}\cdot(\vec{F}_{1}-\vec{v}_{1} \cdot \vec{F}_{1} \vec{\beta})]= \nonumber\\ 
&\epsilon_{2}[1-\vec{v}_{2} \cdot \vec{\beta}-\frac{\vec{r}_{2}-\vec{r}_{1}}{E_{1}+E_{2}}\cdot(\vec{F}_{2}-\vec{v}_{2} \cdot \vec{F}_{2} \vec{\beta})],
\end{align}
where $\vec{F}_{i}=\vec{F}_{s,i}+\vec{F}_{ext,i}$. $\vec{F}_{s,i}$ stands for the infracolor force, and $\vec{F}_{ext,i}$ represents external force of the $i$th quirk. $\vec{v}_{i}$ is the velocity of quirk in the lab frame.

Then, for two quirks at any $(t'_i, \vec{r}'_i)$ according with the relation of Eq.~(\ref{eq::synchrone}), $\hat{s}_{1}$ and $\hat{s}_{2}$ in the lab frame are given by the unit vectors of 
\begin{align}
\vec{r}_{s1}&=(\vec{r}'_1-\vec{r}'_2)-(t'_1-t'_2)\vec{v}'_1~,\label{eq::s1}\\
\vec{r}_{s2}&=(\vec{r}'_2-\vec{r}'_1)-(t'_2-t'_1)\vec{v}'_2~.\label{eq::s2}
\end{align}

To control the truncation error in the Euler's Method, the time step sizes of two quirks should satisfy
\begin{align}
\epsilon_{1,2} < 10^{-4}~\text{ns} \frac{m_{\mathcal{Q}}}{100~\text{GeV}} \frac{1~\text{keV}^2}{\Lambda^2}. 
\end{align}
Finally, the time evolution stops when any of the following criteria are met
\begin{itemize}
\item the evolution time of both quirks is longer than 25 ns, or
\item both quirks are propagating outside the HCal. 
\end{itemize}

\subsection{Thickness of the quirk hits plane}

\begin{figure*}[t]
\begin{center}
\includegraphics[width=1.0\textwidth]{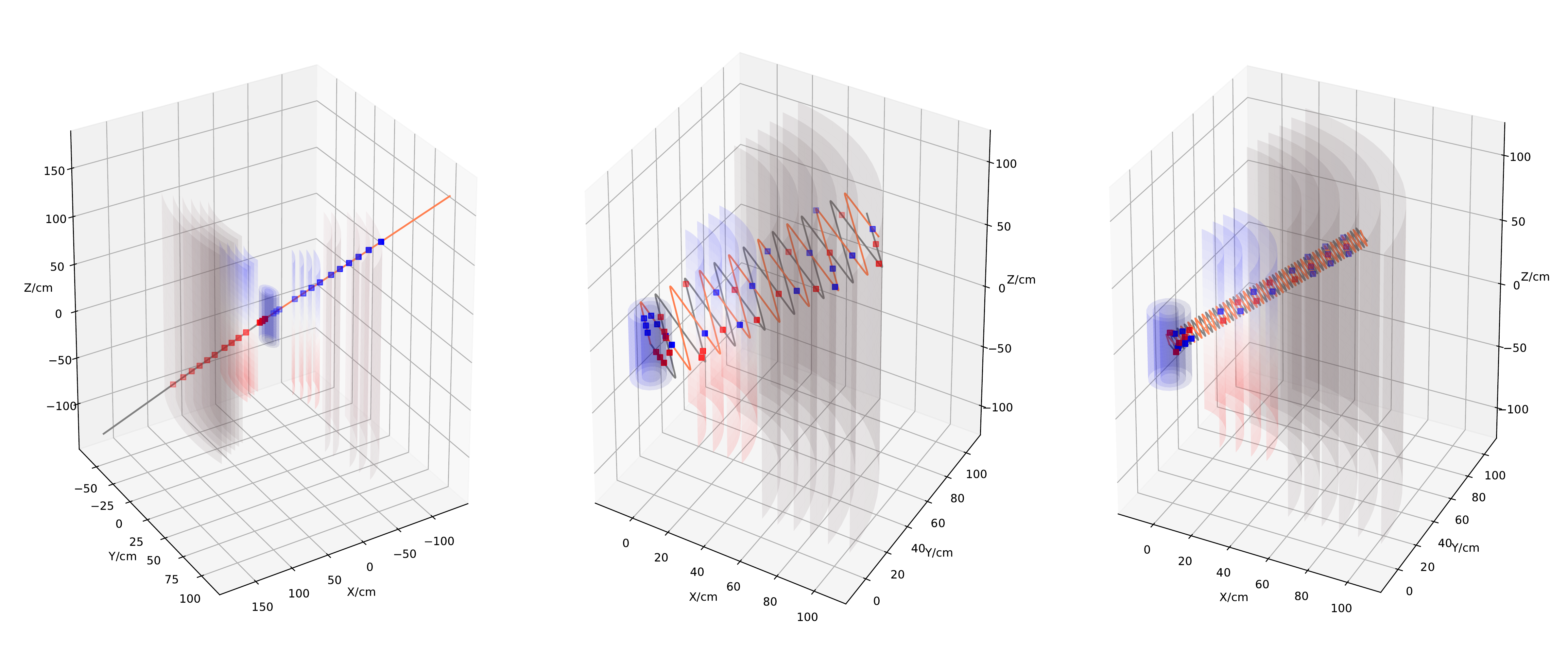}
\end{center}
\caption{\label{fig:track}{Trajectories of two quirks inside the CMS tracker, with the confinement scale chosen as 1 eV, 1 keV and 2 keV (from left to right). The quirk system in different panels have common initial momentum and the quirk mass is 100 GeV. The cylinder segments indicate the barrel layers. } }
\end{figure*}

The CMS tracker is segmented into cylindrical barrels, which surround the beam pipe, and end caps on both sides of the barrels. It consists of two subsystems: the pixel detector and the strip tracker. The details of the CMS detector parameters can be found in Ref.~\cite{Chatrchyan:2008aa}, and we illustrate the barrel layers in Fig.~\ref{fig:track}.  
In our simulation, the resolution of the hit positions in the tracker will be taken as 10 $\mu$m. 

\begin{figure}[thb]
\begin{center}
\includegraphics[width=0.45\textwidth]{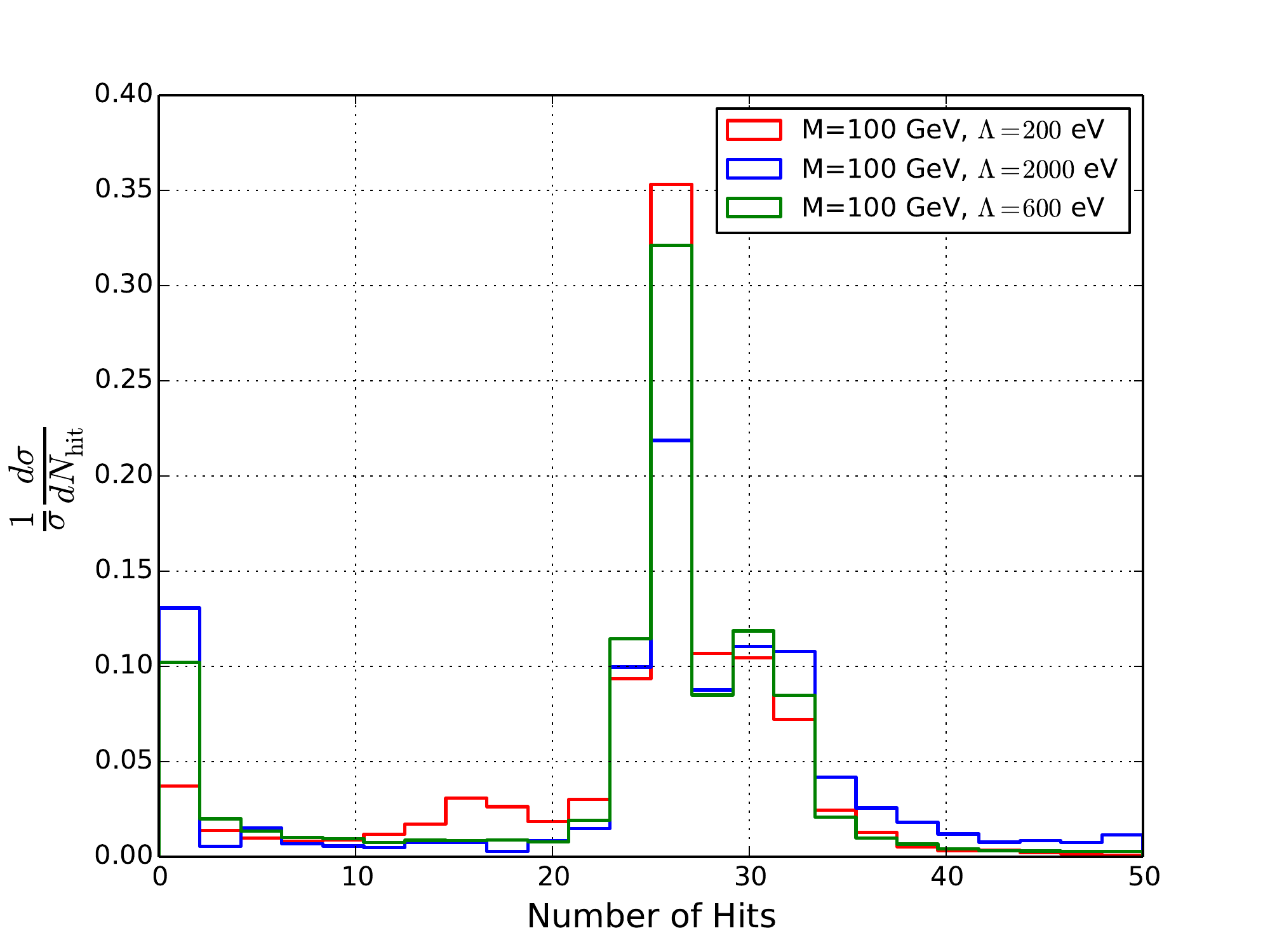}
\includegraphics[width=0.45\textwidth]{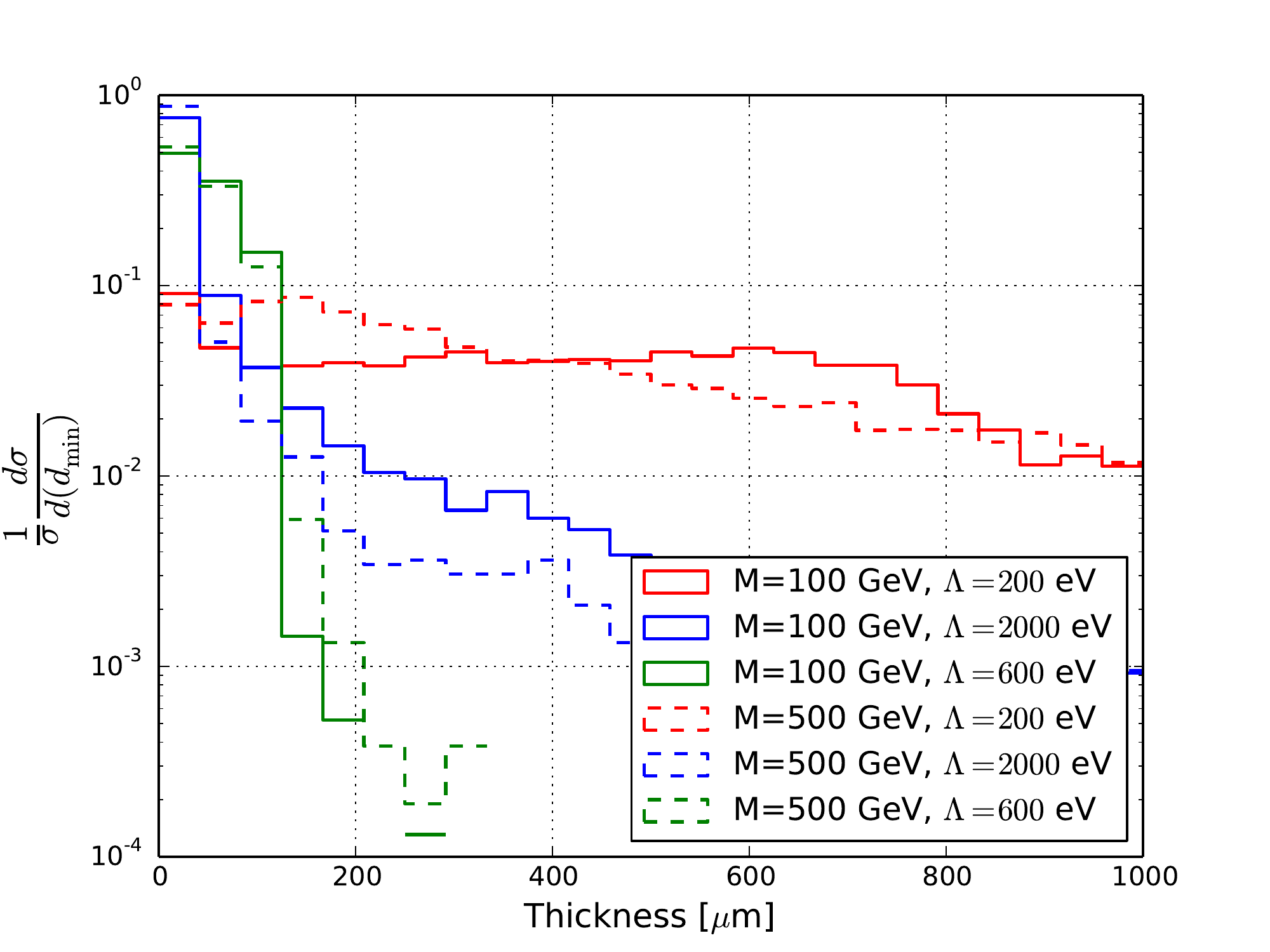}
\end{center}
\caption{\label{fig:hits} The distributions of the total number of hits for the quirk system traveling through the CMS tracker (upper). The distributions of the reconstructed thickness of the quirk pair plane (lower).}
\end{figure}

The charged quirks will leave a number of hits in the tracker, from which we can reconstruct the thickness of the quirk pair plane. In the upper panel in Fig.~\ref{fig:hits}, we plot the distributions for the number of hits of the quirk pair with different confinement scales $\Lambda=200~\text{eV},~600~\text{eV}$, and $2~\text{keV}$. 
In the figure, we have chosen the quirk mass $m_{\mathcal{Q}}=100$ GeV and the transverse momentum of quirk pair $p_T>100$ GeV. We can find that the number of hits is weakly related to the confinement scale. Most of the events have 26 hits in the tracker because the quirk pair with large transverse momenta can go through all 13 barrel layers. 
There is also a great possibility that the quirk system leaves more than 26 hits inside tracker, when the quirk travels into end cap layers or the quirk crosses the same layer more than once. 

Given a set of hit positions $\vec{h}_i$ ($i=1,2,\dots,N$), we can define
\begin{align}\label{eq::thickness}
d(\vec{n})=\sqrt{\frac{1}{N-1}\sum_{i=1}^{N}(\vec{n}\cdot\vec{h}_i)^2}
\end{align}
to describe the average distance between the hits and a virtual plane $\vec{n}$, which also includes the primary vertex.
The $\vec{n}$ that gives the smallest $d(\vec{n})$ (denoted by  $d_{\min}$) corresponds to the normal vector of the quirk plane.
The thickness of the quirk plane is induced by the $\vec{n}$-component of the Lorentz force, so it can be estimated as~\cite{Li:2020aoq}
\begin{align}\label{eq:thickp}
 d_{\min}\sim 1.16 \left(\frac{m}{100~\text{GeV}} \right) \left(\frac{\text{keV}}{\Lambda} \right)^4\left(\frac{|q|}{e} \right)\left(\frac{B_{xy}}{T} \right)\mu\text{m}, 
\end{align}  
where $B_{xy}=|\vec{B}-(\vec{B}\cdot \vec{n}) \vec{n}|$.

The thickness of the quirk plane can also be calculated by the algorithm proposed in Ref.~\cite{Knapen:2017kly}, which is simplified to an eigenvalue problem. We have verified that the thickness calculated from this method matches well with our estimation in Eq.~(\ref{eq:thickp}). 
In the lower panel in Fig.~\ref{fig:hits}, we plot the distributions of the quirk plane thickness, varying both the quirk mass and the confinement scale. 
As before, the quirk pair is required to have $p_T>100$ GeV in the event simulation.
It is clear that the dependence on the confinement scale is much stronger than the dependence on quirk mass. 
We can also observe that for $\Lambda \in [\mathcal{O}(100)~\text{eV},\mathcal{O}(1)~\text{keV}]$, most of the events have a quirk plane thickness smaller than $\sim \mathcal{O}(100)~ \mu$m. In what follows, we will focus on the case with confinement scale $\sim$ keV. \footnote{Even though our quirk signal selection algorithm is applicable to a wide range of $\Lambda$, some parameters in the algorithm discussed below are optimized on the case with $\Lambda \sim 1$ keV. }


\section{Signal and background analyses at the 13 TeV LHC}
\label{sec:ana}

In the Monte Carlo simulation of event samples, the simplified models for four different quirks are implemented in FeynRules~\cite{Alloul:2013bka}. The general purpose event generation framework MG5\_aMC@NLO~\cite{Alwall:2014hca} is used to simulate the quirk production processes with the model file provided by FeynRules. Then the built-in Pythia8~\cite{Sjostrand:2007gs} is used for implementing the parton shower, hadronization and decay of the SM particles.  
However, the QCD parton shower as well as the hadronization of the colored quirk is neglected for simplicity. 

Because the missing transverse energy trigger is adopted in the quirk search, the dominant background is the $Z(\to \nu\nu)+$jets process. In generating the $Z(\to \nu\nu)+$jets events, we require the transverse momentum of the $Z$ boson to be $p_T(Z)>200$ GeV. This is conservative, since a much stronger cut will be applied in the later analysis. The MLM prescription~\cite{Hoche:2006ph} is used for matching of matrix element (up to two jets) with parton showers. 
The events with exactly one jet in the final state are selected as in the monojet search.
We also consider the background process of $Z(\to \nu\nu) e^+ e^- +$ jet as pointed out in Ref.~\cite{Knapen:2017kly}, with $p_T(Z)>100$ GeV  and $p_T(e) > 1$ GeV.  The production cross sections for $Z(\to \nu\nu)+$jet and $Z(\to \nu\nu) e^+ e^- +$ jet after applying preliminary $p_T$ cuts are 3.6 pb and 13.95 fb, respectively. 
Note that the $W(\to \ell \nu)$+jet process can have relatively large missing transverse energy as well. However, this process can be suppressed by vetoing events that contain either an electron or muon or a reconstructed $\tau$. Only $W(\to \ell \nu)$+jet events with missing lepton and $W(\to \tau \nu)$+jet events with the hadronically decaying $\tau$ being mis-tagged as a QCD jet can survive. Its rate is much smaller than the rate of $Z(\to \nu\nu)+$jets event, so we will not consider this background. 

The trajectory of a charged SM particle in the tracker is a helix, which is determined by the initial momentum of the particle. The hits of the particle can be obtained by calculating the intersecting points between the helix and the tracking layers. 
As for quirks, after solving the EOM numerically as discussed in Sec.~\ref{sec:motion}, we illustrate their trajectories inside the CMS tracker in Fig.~\ref{fig:track}. 
In all three panels, the quirk masses and initial momenta are fixed to the same values, while the confinement scales are chosen to be 1 eV, 1 keV and 2 keV, respectively. 



\begin{figure}[thb]
\begin{center}
\includegraphics[width=0.45\textwidth]{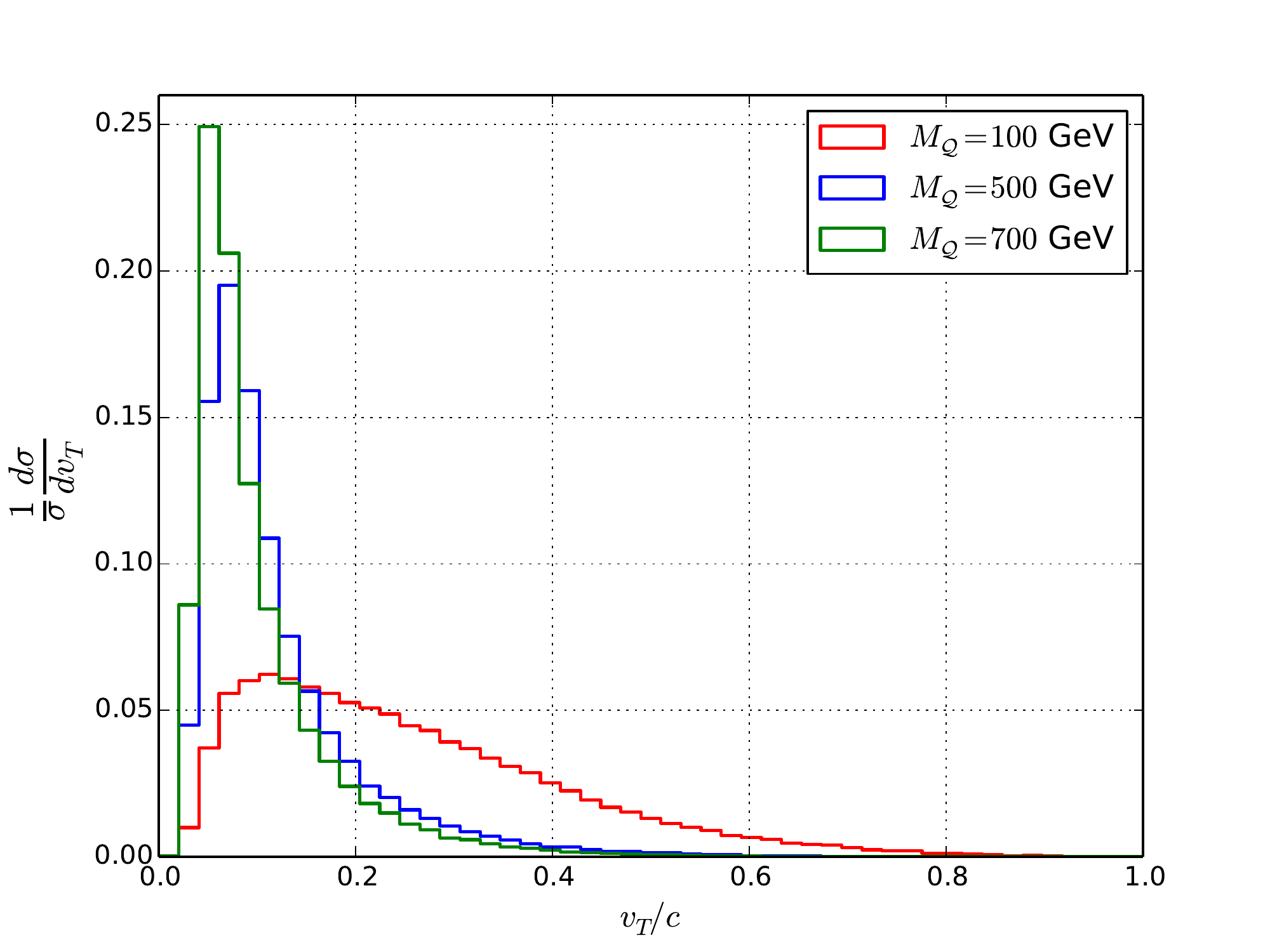}
\includegraphics[width=0.45\textwidth]{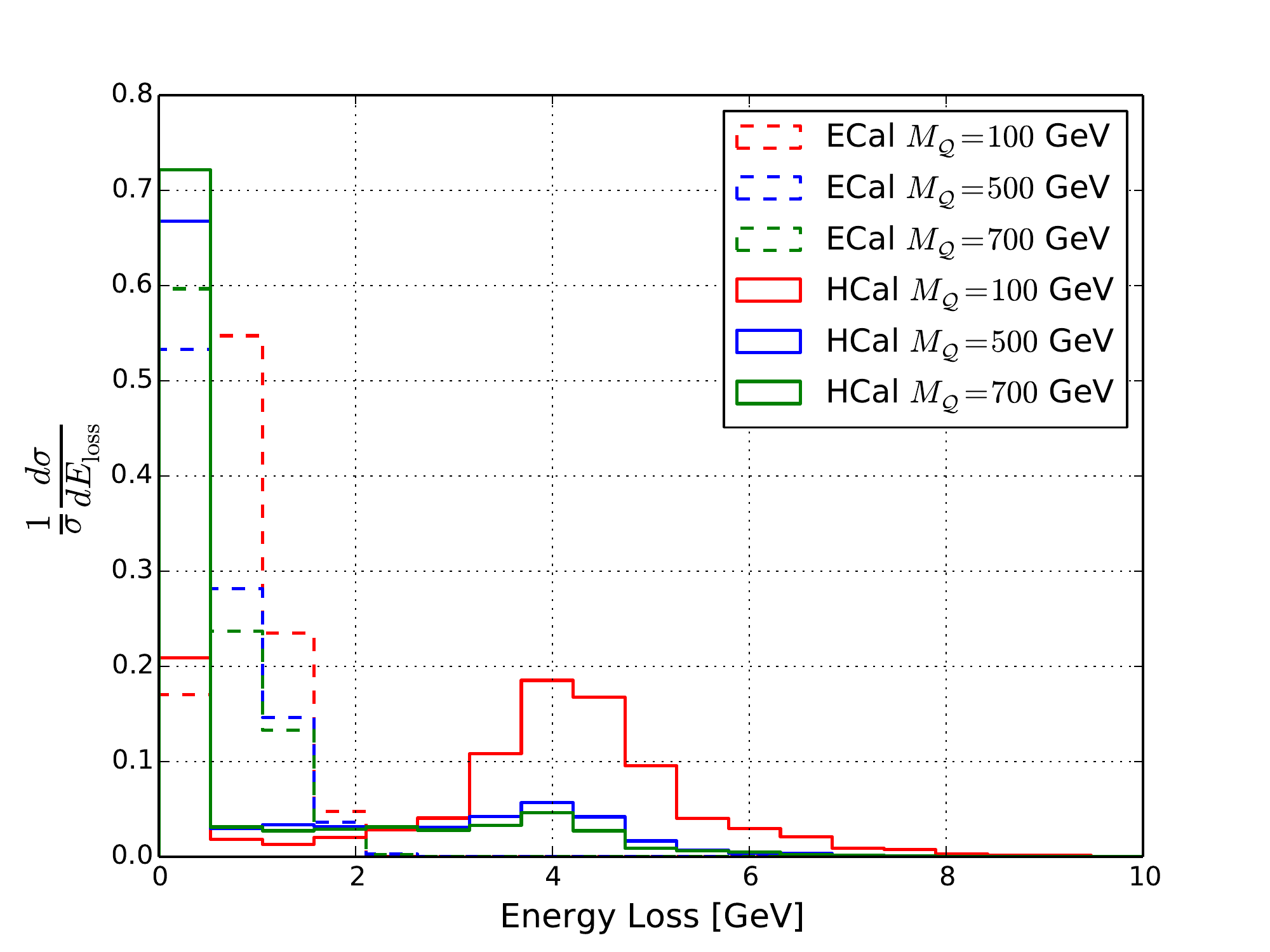}
\end{center}
\caption{\label{fig:vtmet} The distributions of transverse velocity of the quirk pair system (upper panel).  The distributions of energy loss of the quirk pair in the ECal and HCal (lower panel). Three different masses of quirk have been chosen for presentation. The quirk pair is required to have $p_T>100$ GeV in the event simulation. }
\end{figure}

Before moving on to designing the quirk search method, we discuss a few features of the quirk's motion in the detector. 
The duration of a quirk traveling inside the detector is proportional to the inverse of transverse velocity $v_T$ of the quirk pair system. The distribution of $v_T$ is shown in the upper panel in Fig.~\ref{fig:vtmet}. 
Even though each quirk can have velocity $\gtrsim 0.5$, the quirk pair system is moving slowly with $v_T \sim 0.1$, especially when the quirk is heavy. 
As a result, the quirk system can escape the tracker within $10-20$ ns (given the CMS detector). 
While the quirk is traveling in the calorimeter, it will lose energy by ionization~\footnote{The energy lost through hadronic interactions with the detector is much smaller than energy lost through electromagnetic ionization ~\cite{Aad:2013gva}. We will not consider this contribution in our simulation.}. According to our simulation, given $\Lambda \sim \mathcal{O}(1)~\text{keV}$, $m_{\mathcal{Q}} \sim \mathcal{O}(100)$ GeV, and transverse momentum of the quirk pair $p_T>100$ GeV, the total energy losses inside the ECal and HCal are around 1 and 4 GeV, respectively, as shown in the lower panel of the Fig.~\ref{fig:vtmet}. It should be noted that we count only the energy deposition within the colliding period of 25 ns. According to our simulation, there are $\mathcal{O}(10)$\% ($\gtrsim 50$\%) of quirks that cannot reach the ECal (HCal) of the CMS detector within 25 ns. Most of the quirk pairs are still propagating in the HCal after 25 ns.




\subsection{Pileup simulation}
During a bunch crossing period at the LHC, there are multiple proton-proton collisions (referred as pileup), which are dominated by the nondiffractive events (the sample is referred as the minimum bias) with small transverse momenta transfer.  Pythia8~\cite{Sjostrand:2007gs} adopts perturbative parton shower, Lund-string hadronization, multiple parton interaction and colour reconnection models to simulate the minimum bias. However, those models contain many parameters, which can only be deducted from experimental data. The set of chosen parameters is referred as Pythia tunes~\cite{ATLAS:2012uec}. The A3 tune with phenomenological parameters provided in Refs.~\cite{Skands:2014pea,ATL-PHYS-PUB-2016-017} is taken in our simulation of pileup events, because it is found to provide a good agreement with the charged particle distributions at the ATLAS detector. 

The number of pileup events per bunch crossing at the LHC follows Poisson distribution with an average value around $\langle \mu \rangle \sim 30-50$. They will give $\mathcal{O}(10^4)$ hits inside the tracker, and, 
thus become the main background in our analysis. 
In our simulation, both signal and background events are overlaid by $\langle \mu \rangle = 50$  pileup events in order to draw a conservative conclusion. Moreover, due to the finite size of the beam spot, the interaction points of the pileup events could be spread along the z direction. We will assume the z coordinate of the interaction points follows a Gaussian distribution with a width of 45 mm. 

The hits caused by pileup events can be obtained by the same method as has been used for the $Z+$jets background, {\it i.e.}, 
intersecting the helix trajectories of the charged particles with the CMS tracking layers. 

\subsection{Quirk signal selections}
\label{selection}
In this section, we will propose a dedicated algorithm to separate the quirk hits from the SM particle hits. 
In the following discussion, the quirk refers to the $\tilde{\mathcal{D}}^c$ and $\Lambda =1$ keV,  even though our algorithm is also applicable to quirks with different quantum numbers and different values of $\Lambda$. 

In searching for the quirk signals, we should first find the plane that quirks move on based on the recorded hits inside the tracker. 

\begin{itemize}
\item[A.]
First, the hits with $dE/dx$ smaller than 3.0 MeV/cm are removed to reduce the number of background hits. The quirk hits are kept intact at this stage. The remaining hits are classified according to the $dE/dx$. 
For each class, the average $dE/dx$ is defined as 
\begin{equation}
\left( \frac{dE}{dx}\right)_{\text{avg}}^a=
\frac{1}{N_a} \Sigma_{i=1}^{i=N_a} \left(\frac{dE}{dx}\right)_i 
\end{equation}
where $N_a$ is the number of hits in the ath class. 
We loop over all the hits. A hit is assigned to the class $a$ if $| \left( \frac{dE}{dx} \right)_{\text{this hit}}-\left(\frac{dE}{dx}\right)_{\text{avg}}^a|<1.0~\text{MeV/cm}$. Otherwise, a new class is created and the hit is put into it. 

After the hit classification, the classes with fewer than 80 hits~\footnote{The number of 80 is chosen such that the hits from quirk are kept as much as possible, while maintaining an affordable time consumption. } will be used directly in step B. 
On the other hand, in classes with more than 80 hits, the sets of hits which belong to helix trajectories are removed. This step is found to be very useful for pileup hit removal.
In practice, we will try to reconstruct the circles in the transverse plane instead of reconstructing the true helixes.
The location of the center of the circle (COC) in the transverse plane for any two hits (plus the origin) is given by
\begin{align}
x_0&=\frac{1}{2\left(k_1-k_2\right)}\left[k_1x_2\left(1+k^2_2\right)-k_2x_1\left(1+k^2_1\right)\right] ~, \nonumber\\
y_0&=-\frac{x_0}{k_1}+\frac{x_1}{2k_1}+\frac{y_1}{2} ~, \nonumber
\end{align}
where $k_1=y_1/x_1$ and $k_2=y_2/x_2$. $x_1,y_1$ and $x_2,y_2$ are $x$-$y$ coordinates of the two hits. 
In each iteration, we pick out $80\times 80$ pairs of hits in each class randomly (even in the class with a number of hits much larger than 80). After calculating the position of the COC and the radius for each hit pair, the list of the selected hit pairs in each class is sorted by the size of the radius~\footnote{The time complexity is $n_{\text{class}} \times \mathcal{O}(n\log 2n) \sim \mathcal{O}(10^6)$.}. Then, we compare the position of the COC and the radii of the adjacent hits. If the distance between the two COCs is smaller than 0.5 cm and the difference between the two radii is smaller than 0.1 cm~\footnote{Considering the the trajectories are a slightly distorted arc of helices due to experimental effects~\cite{Amrouche:2019wmx}. }, the two pairs of hits are considered to be induced by the same SM particle. 
Subsequently, all of the hits that lie on the circle will be removed from the class. The iteration continues until the total number of hits in all classes is less than 1500. 

\begin{table}[htbp]\centering
 \begin{tabular}{|c|c|c|c|c|c|c|} \hline
            & $ \langle \epsilon_1 \rangle$ & $\langle \epsilon_2 \rangle$ & $\langle \epsilon_3 \rangle$  & $\langle \epsilon_4 \rangle$  &   $\langle \epsilon^{1500} \rangle$    \\ \hline
  $m_{\mathcal{Q}}=200~\text{GeV}$          & 0.965 & 0.590 & 0.5010 & 0.443  & 0.903             \\ \hline
  $m_{\mathcal{Q}}=500~\text{GeV}$          & 0.933 & 0.356 & 0.307  & 0.278  & 0.899             \\ \hline
  $m_{\mathcal{Q}}=800~\text{GeV}$          & 0.910 & 0.372 & 0.316  & 0.276  & 0.871             \\ \hline
  Background                 & 0.167 & 0.064 & 0.055  & 0.049  & 0.153             \\ \hline
  \end{tabular}
  \caption{\label{tab:reduce} The average signal and background reduction efficiencies  at the $i$th iteration. The last column is the average reduction efficiencies when the remaining number of hits goes below 1500. }
\end{table}

In Table~\ref{tab:reduce}, we list the average reduction efficiencies after $i$th iterations $\langle \epsilon_i \rangle$ and the average reduction efficiencies when the remaining number of hits is less than 1500 $\langle \epsilon^{1500} \rangle$. The reduction efficiencies for the quirk and background are defined as the ratio between the number of quirk(background) hits after the removal and the total number of quirk(background) hits. The $\langle \epsilon_i \rangle$ is obtained by averaging all events. 
We can see that one iteration can already reduce the number of hits to close to 1500 in both quirk and background events. 

\item[B.]
Second, we should calculate the normal vector of the virtual quirk plane based on the hits after the selections in step A, which include all the hits in hit classes with fewer than 80 hits and the remaining hits in the hit classes with 80 hits or more. 
Assuming the origin is contained in the quirk plane, the plane normal vector can be determined for any two hits:
$\vec{n}=\langle\vec{r}_1\times\vec{r}_2\rangle$, where $\vec{r}_1$ and $\vec{r}_2$ are the coordinates of the two hits. 
For a given plane, the distance of a new hit ($\vec{r}_i$) to the plane is  $d_i=\vert\vec{r}_i\times\vec{n}\vert$. 
Our choice of $\Lambda=1$ keV leads to a typical plane thickness of $\sim \mathcal{O}(10^{-5})$ m. So we count the number of hits that are within a distance of $30~\mu$m to the plane. The plane~\footnote{Since there are $\sim 20$ quirk hits out of 1500 hits, $\mathcal{O}(1000)$ tries are already able to find the quirk plane. Here we choose to scan over $10^5$ randomly selected hit pairs.} with the largest number of counted hits is kept and will be regarded as the quirk plane. 
In the upper panel in Fig.~\ref{fig:sp}, we plot the distributions of the number of hits ($N_{\text{left}}$) within the quirk plane ($d<30~\mu$m) for signals with different quirk masses as well as the background. We can observe that the signals tend to have much larger number of hits than the background. 
Moreover, the cutoff at $N_{\text{left}}=26$ is attributed to the fact that two quirks go through all 13 barrel layers of the CMS tracker, leaving 26 hits in total. 

\begin{figure}[t]
\begin{center}
\includegraphics[width=0.4\textwidth]{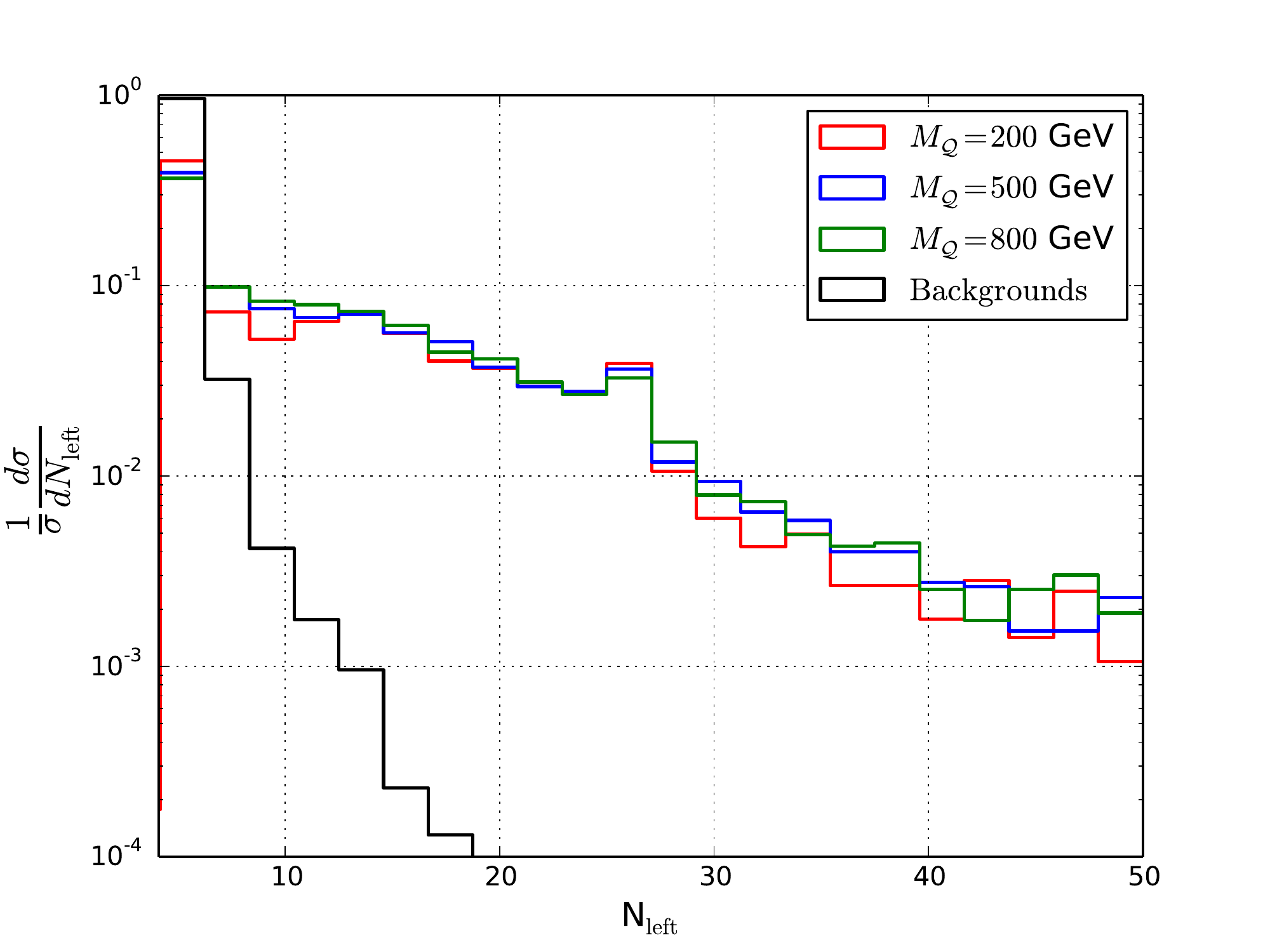}
\includegraphics[width=0.4\textwidth]{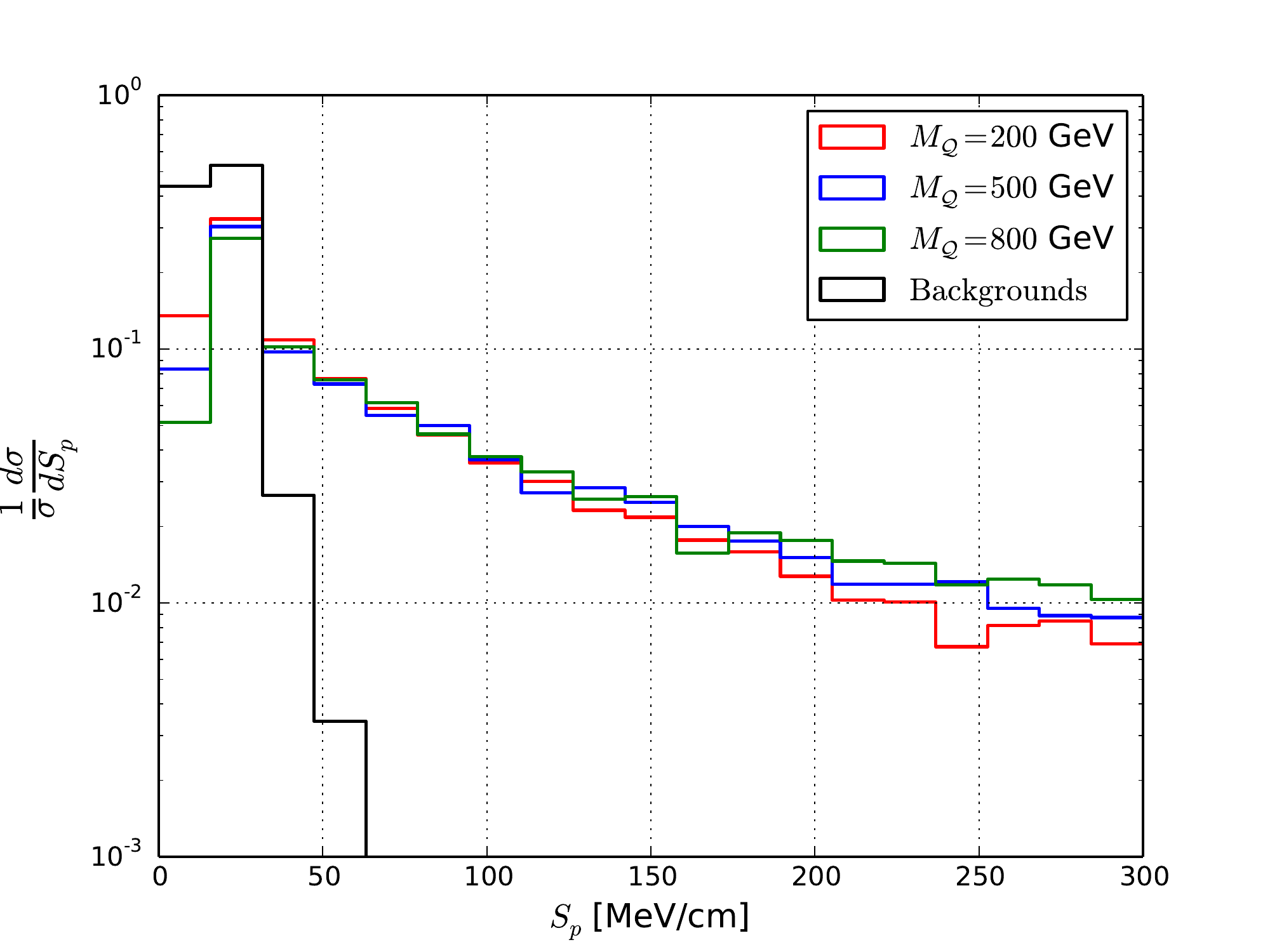}
\end{center}
\caption{\label{fig:sp} The distributions for the number of hits within $d<30~\mu$m of the quirk plane (upper), and the distance-weighted ionization energy loss $S_p$ (lower).  }
\end{figure}

\item[C.]
Third, based on the reconstructed quirk plane (with normal vector $\vec{n}_p$), we can define two important variables for further signal and background discrimination.
The distance-weighted ionization energy loss is defined as
\begin{equation}
S_p=\sum_{i} e^{-d_i/d_0} \times \left(\frac{dE}{dx}\right)_i, \label{eq:sp1}
\end{equation}
where $i$ runs over all remaining hits after step A and $d_0=30~\mu$m. 
Furthermore, the variation of the $S_p$ 
\begin{equation}
\Delta_p= \max_{\Delta \theta(\vec{n}_p, \vec{n}_{p'})<\pi/12}\frac{|S_p-S_{p\prime}|}{S_p}
\end{equation}
can also be useful for signal identification. Here, the $S_{p\prime}$ is also calculated by Eq.(\ref{eq:sp1}), but the direction of the reference plane is varying within an angle of $\frac{\pi}{12}$ around $\vec{n}_p$. 
The distributions of $S_p$ for signals and background are shown in the lower panel in Fig.~\ref{fig:sp}. 
Because of the large ionization energy loss of the quirk, the signals have significantly harder $S_p$ spectra than the background. 

\end{itemize}
 
Finally, as we have discussed in Sec.~\ref{sec:ana}, the large missing transverse energy ($E^{\text{miss}}_T$) is also an important feature of the quirk signal, so we propose the following cuts for the quirk search at the LHC:
\begin{itemize}
\item[1.] $E^{\text{miss}}_T$$\geq$ 500 GeV,
\item[2.] $\Delta_p \geq 0.8$,
\item[3.] $S_p \geq 50$ MeV/cm, and
\item[4.] $N_{\text{left}}\geq 10$. 
\end{itemize} 
The cut flow of several quirk signals and backgrounds are given in Table~\ref{tab:cuts}, where we can find that the $E^{\text{miss}}_T$ cut and $S_p$ cut play the most important roles.~\footnote{The lepton veto does not affect the results}.

\begin{table}[htbp]\centering
 \begin{tabular}{|c|c|c|c|c|c|c|} \hline
 \multirow{2}{*}{} & \multicolumn{3}{c|}{Quirk mass} & \multirow{2}{*}{$Z e e j$ } & \multirow{2}{*}{$Z+$jets} \\ \cline{2-4}
  
                   & 200 GeV   & 400 GeV    & 800 GeV   &    &          \\ \hline\hline
  $\sigma$ [fb]                        & 33230  & 1484   & 29.92  & 13.95      & 3600\\ \hline
  $E^{\text{miss}}_T \geq 500$ GeV     & 984    & 76.7    & 2.20   & $0.02$     & 84.5\\ \hline
  $\Delta_p \geq 0.8$                  & 910    & 72.6    & 2.09   & $0.018$    & 72.5 \\ \hline
  $S_p \geq 50$ MeV/cm                 & 429    & 33.6    & 1.21   & $\sim 0$   & $0.51$ \\ \hline
  $N_{\text{left}}\geq 10$             & 413    & 28.9    & 1.01   & $\sim 0$   & $0.12$ \\ \hline
  \end{tabular}
  \caption{\label{tab:cuts} Cut flows of our analysis for the colored scalar quirk $\tilde{\mathcal{D}}^c$ pair productions and for the SM background processes. Three different masses of quirk have been chosen for illustration.  The numbers in the table correspond to the cross sections (in femtobarns) at the 13 TeV LHC.  }
\end{table}

\begin{figure}[t!]
\begin{center}
\includegraphics[width=0.45\textwidth]{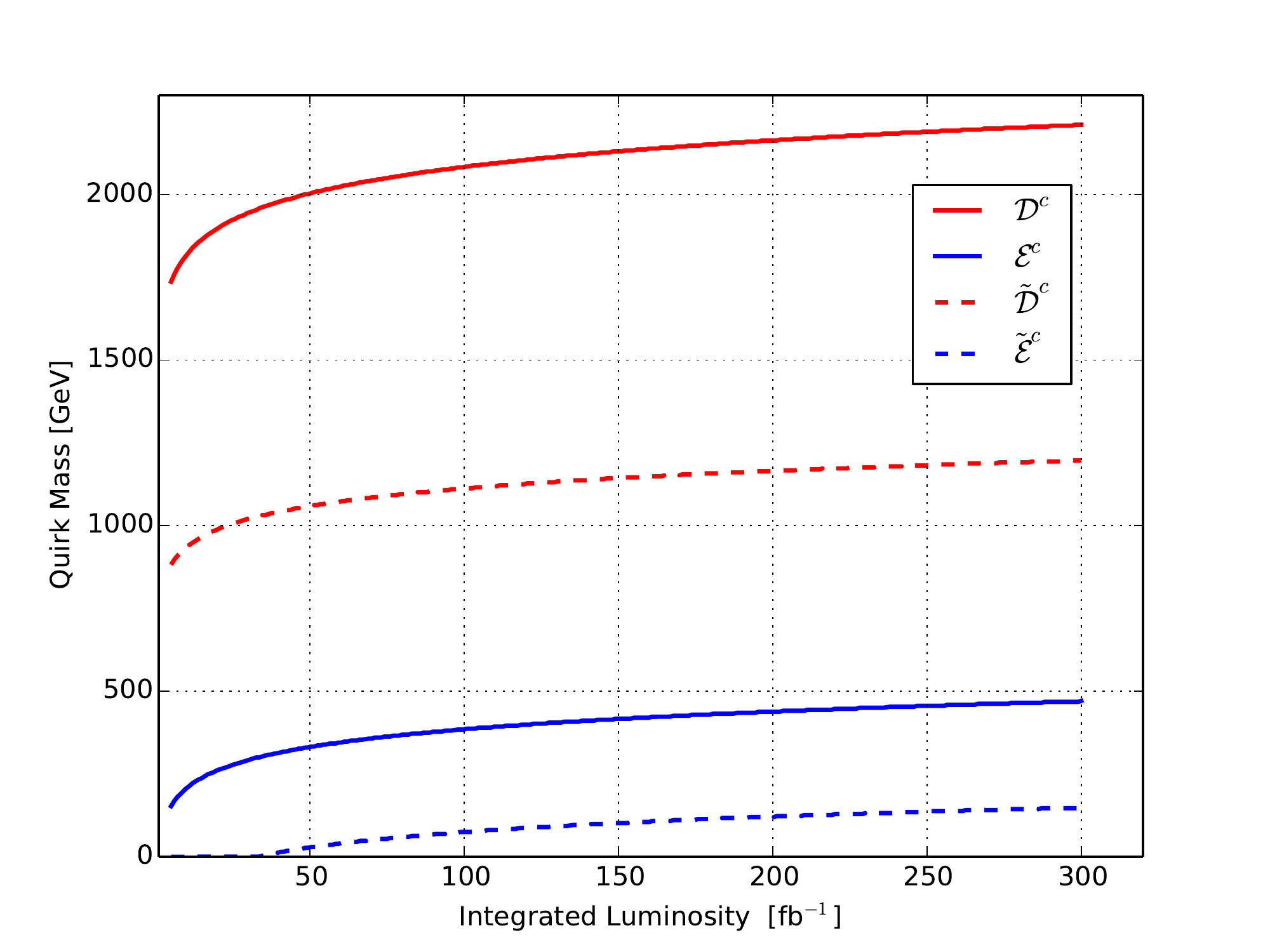}
\end{center}
\caption{\label{fig:exc} The 95\% C.L. exclusion limits for quirks with different quantum numbers at 13 TeV LHC.  }
\end{figure}

In Fig.~\ref{fig:exc}, we plot the 95\% confidence level (C.L.) exclusion limits for quirks with different quantum numbers at 13 TeV LHC. 
The search sensitivities for quirks with different quantum numbers are significantly different, attributed to two facts. Most importantly, the production cross sections of quirks are quite different as shown in Fig.~\ref{fig:xsec}. Moreover, quirks with different quantum numbers have different production channels, which lead to different quirk momentum distributions. 
The integrated luminosity at the LHC run 2 is $\sim 100$ fb$^{-1}$. It means  the colored fermion(scalar) quirks with masses up to {2.1(1.1) TeV}, and color neutral fermion(scalar) quirks with masses up to {450(150) GeV} can be possibly probed or excluded. 

\section{Conclusion}
\label{sec:conc}
The quirk widely exists in new physics models with extra gauge symmetries. The color neutral quirk is well motivated by the neutral naturalness model which can solve the little hierarchy problem. We consider a simplified model framework for quirks with different quantum numbers and study their discovery prospects at the LHC. 

The quirk equation of motion inside the detector is mainly controlled by the Lorentz force and the long-range infracolor force as well as the frictional force from ionization energy loss. In quirk pair production, the infracolor forces between two quirks are correlated, leading to a coupled system of nonlinear partial differential equations. It can be solved numerically in the limit that the infracolor force is much larger than the other external forces, such that the string connecting two quirks is approximately straight in the quirk pair center of mass frame. 
We numerically solved the full EOM for the quirk system, taking into account the architecture of the CMS detector. 
For the parameter space of interest [{\it i.e.} $\Lambda \sim \mathcal{O}(100-1000)$ eV, $m_{\mathcal{Q}} \sim \mathcal{O}(100)$ GeV and $p_T(\mathcal{Q}\bar{\mathcal{Q}}) \gtrsim 100$ GeV], most of the quirk pairs can leave total number of $\sim 26$ hits inside the tracker. Those hits lie on the plane with thickness $\lesssim \mathcal{O}(100) ~\mu$m. 
Meanwhile, there will be $\mathcal{O}(10)$\% ($\gtrsim 50$\%) of quirks that cannot reach the ECal (HCal) of the CMS detector within 25 ns (the LHC bunch spacing). 
The total energy losses of quirks in both the ECal and HCal within this time interval are found to be small ($\lesssim 5$ GeV). 

The main background in the quirk search is the SM $Z(\to\nu\nu)$+jets process overlaid by abundant pileup events (we take $\langle \mu \rangle =50$).  We propose a dedicated hit reduction algorithm to remove the SM particle hits while keeping the quirk hits as much as possible. The quirk plane is reconstructed for each event based on the selected hits. Subsequently, three discriminative variables can be defined: the number of remaining hits within a distance of  $30~\mu$m to the quirk plane, the distance-weighted ionization energy loss in the tracker, and the variation of the $S_p$. 
Together with the missing transverse energy, they can provide a very sensitive probe for the quirk signal. 
With the parameters in the search optimized on the case with $\Lambda\sim 1$ keV, we find that the $\sim 100$ fb$^{-1}$ dataset at the LHC will be able to probe the colored fermion(scalar) quirks with masses up to {2.1(1.1) TeV}, and color neutral fermion/scalar quirks with masses up to {450(150) GeV}, respectively.  

Compared to the results in Ref.~\cite{Knapen:2017kly}, where it predicts that the 300 fb$^{-1}$ dataset is able to probe the $\mathcal{D}^c$ (with $N_{\text{IC}}=2$) lighter than 1.5 TeV and the $\mathcal{E}^c$ (with $N_{\text{IC}}=2$) lighter than 500 GeV, the corresponding bounds for the two quirks in our analysis are 2.1 TeV and 400 GeV, respectively. 
There is a substantial improvement of the search sensitivity in the heavy quirk region ($m_{\mathcal{Q}} \gtrsim 1$ TeV). On the other hand, due to the stringent $E^{\text{miss}}_T$ cut in our analysis, the sensitivity in the light quirk region is reduced. Some cuts in our analysis should be adjusted accordingly, if one aims to search for a color neutral quirk. 
Moreover, the $Z(\to\nu\nu)+$jets background is not considered in Ref.~\cite{Knapen:2017kly}, which leads to overoptimistic results.  
We reproduce their analysis~\footnote{In Ref.~\cite{Knapen:2017kly}, they assume 8 layers for the ATLAS tracker. We use the actual ATLAS tracker which has 3 layer for the pixel detector and 4 layers for the silicon-strip detector.}  and apply it to the $Z(\to\nu\nu)+$jets (overlaid by pileup events) background.  
We find that the efficiency for reconstructing the quirk plane that contains at least one hit in each layer in the background events is $\sim 10^{-4}$. This value is close to the efficiency of pileup background as given in Ref.~\cite{Knapen:2017kly}.
In fact, this efficiency is dominated by the probability of finding five coplanar hits in three outer layers [up to step 2(a) in the reference], which is $\sim 10^{-3}$. 
Background events with five such coplanar hits usually include two charged particles flying in a similar direction. It will not be difficult to have their hits lying on a narrow strip if these two particles are energetic.  
Following their analysis, we find the efficiency of requiring ``all but one layer must contain two hits'' is $\sim 10^{-2}$. 
After the final selection, the cross section of $Z(\to\nu\nu)+$jets (overlaid by pileup events) is $\sim 10^{-2}$ fb, which cannot be neglected. 
A more detailed comparison of the sensitivities of two methods is given in the Appendix~\ref{app:comp}.

Finally, let us give more comments on the ionization energy loss of quirks in the tracker. 
Each quirk accelerates and decelerates along its trajectory through the detector, while the quirk pair system is moving steadily with $\beta\sim 0.1$. During the oscillation, the quirk takes longer time in slower state. So there is a greater probability that the quirk crosses the tracking layers with smaller velocity, thus inducing higher ionization energy loss. 
It should be noted that the pattern of the ionization energy loss in different layers could be helpful for further signal and background separation~\cite{Patton:2018vcl} as well as quirk property characterization. We leave this for future work.

\begin{acknowledgments}
This work was supported in part by the Fundamental Research Funds for the Central Universities, 
 by the NSFC under grant No. 11905149,  by the Projects 11847612 and 11875062 supported by the 
National Natural Science Foundation of China, and by the Key Research Program of Frontier Science, CAS.

\end{acknowledgments}

\appendix
\section{Comparison of sensitivities} \label{app:comp}
To make a comparison between the sensitivity of our method and the one proposed in Ref.~\cite{Knapen:2017kly}, we apply both methods to two fermionic quirks $\mathcal{D}^c$ and $\mathcal{E}^c$ , with $N_{\text{IC}}=2$ (same as in the Ref.~\cite{Knapen:2017kly}). In reproducing the analysis of the Ref.~\cite{Knapen:2017kly}, the fiducial efficiency and the reconstruction efficiency for quirk signals are taken as 0.28 and 0.1 for simplicity. We will calculate the leading order production cross section of quirk signal after the trigger ($p_T>200$ GeV) by using MG5\_aMC@NLO. The cross section of the $Z(\to\nu\nu)+$jets background after all of the selections is taken as $10^{-2}$ fb. 

\begin{figure}[h]
\begin{center}
\includegraphics[width=0.45\textwidth]{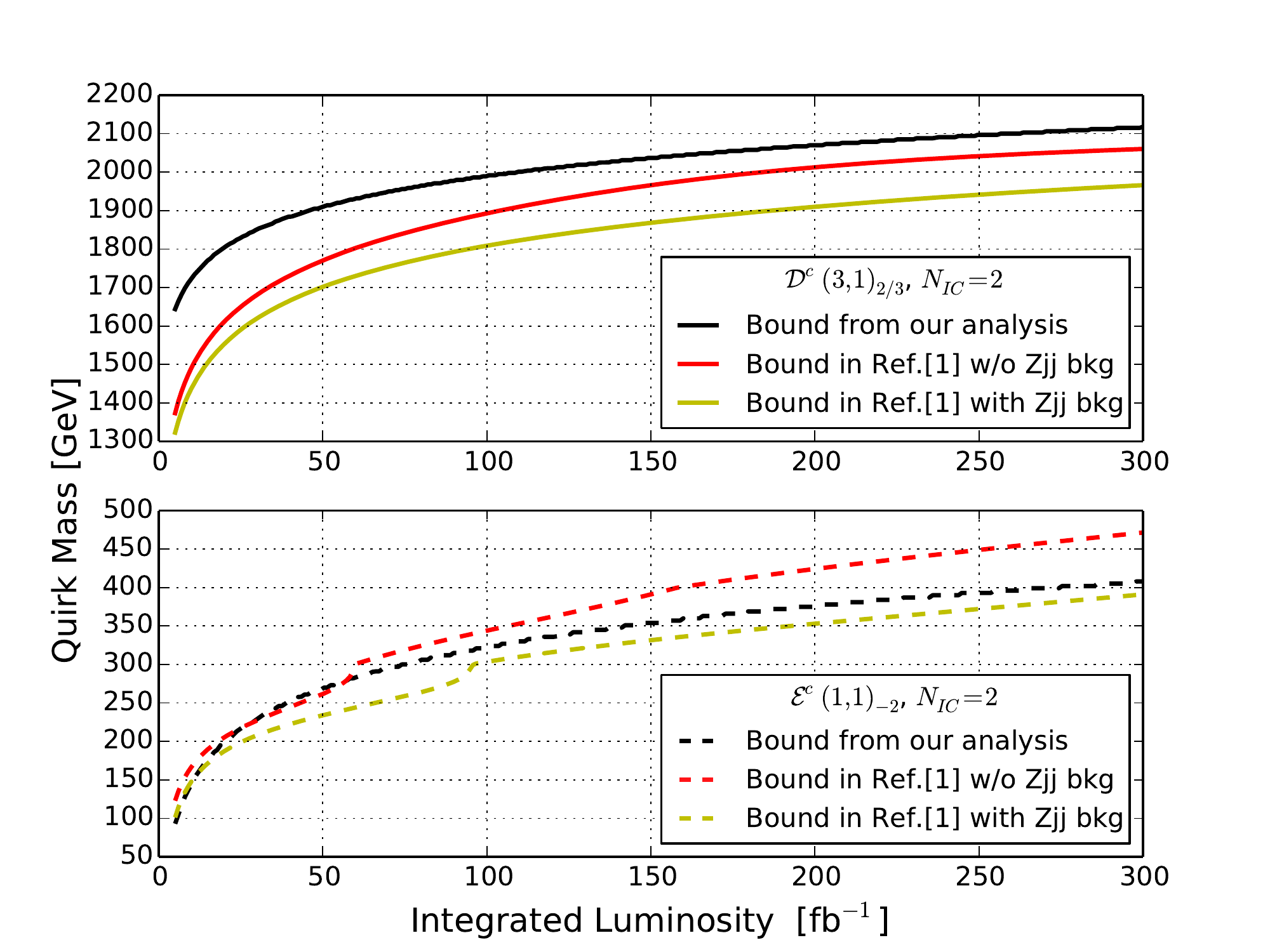}
\includegraphics[width=0.45\textwidth]{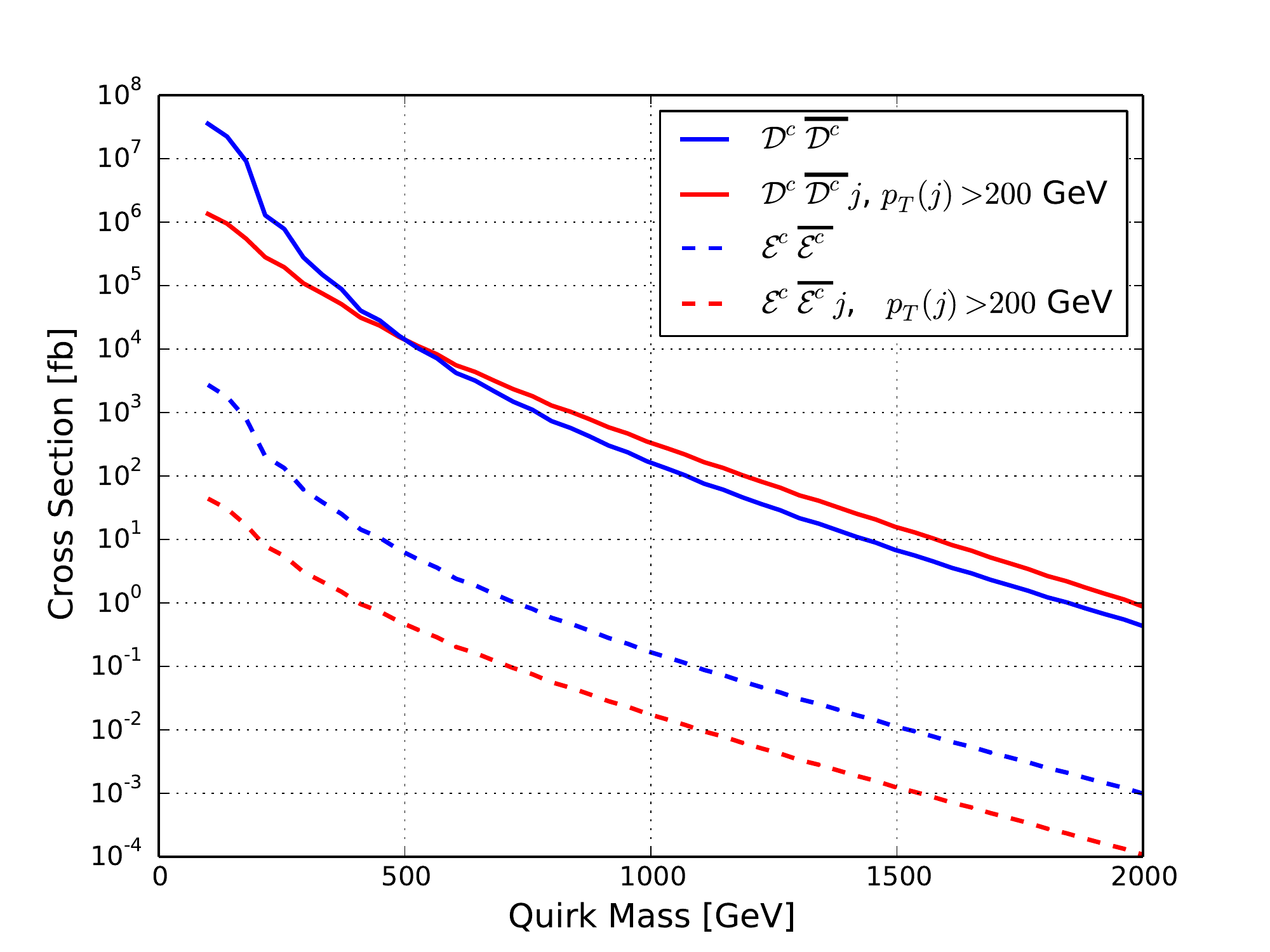}
\end{center}
\caption{\label{fig:comp} Upper: The 95\% confident level exclusion bounds for two different fermionic quirks, from our analysis (black line), from the analysis in the Ref.[1] with (yellow line) and without (red line) considering the $Zjj$ background. Lower: the leading order production cross sections of quirk pair with and without associating jet at the 13 TeV LHC. The jet is required to have $p_T>$200 GeV. }
\end{figure}

The 95\% C.L exclusion limits for $\mathcal{D}^c$ and $\mathcal{E}^c$ obtained from three different methods are shown in the upper panel of the Fig.~\ref{fig:comp}: 
\begin{itemize}
\item Our method with all of the backgrounds (black lines).
\item The method proposed in Ref.~\cite{Knapen:2017kly} without considering the $Z(\to\nu\nu)+$jets background (red lines, they are presented for validation). 
\item The method proposed in Ref.~\cite{Knapen:2017kly} with the  $Z(\to\nu\nu)+$jets background (yellow lines).
\end{itemize}
When the background of $Z(\to\nu\nu)+$jets is considered, our method shows better sensitivities for both $\mathcal{D}^c$  and $\mathcal{E}^c$:  the limits are reinforced by about 100-200 GeV and $\mathcal{O}(10)$ GeV for  $\mathcal{D}^c$  and $\mathcal{E}^c$, respectively. 

There are two points that need to be clarified: (1) our result for the $\mathcal{E}^c$ quirk coincides well  with that in the Ref.~\cite{Knapen:2017kly}, i.e. without considering the $Zjj$ background, 300 fb$^{-1}$ dataset can exclude the $\mathcal{E}^c$ quirk with mass below 500 GeV. (2) the bound of $\mathcal{D}^c$ which we reproduce is much stronger than that in the Ref.~\cite{Knapen:2017kly}, i.e. at 300 fb$^{-1}$, the bound is 1500 GeV in the Ref.~\cite{Knapen:2017kly}, while it is 2050 GeV in our simulation (without considering the $Zjj$ background).  This deviation may be attributed to the incorrect cross section that is used in the Ref.~\cite{Knapen:2017kly} for the $\mathcal{D}^c$. In the lower panel of the Fig.~\ref{fig:comp}, we provide the leading order cross section (calculated by MG5) for quirks either with or without associating jet. We can find that the trigger efficiency ($p_T(j)>$200 GeV) for $\mathcal{E}^c$ production process is 0.1, which coincides with the number in the Table I of the Ref.~\cite{Knapen:2017kly}. However, for $\mathcal{D}^c$, Ref.~\cite{Knapen:2017kly} finds the trigger efficiency is 0.24 and our result shows that the cross section of $p p \to  \mathcal{D}^c \mathcal{D}^c j $ is even larger than that of $p p \to  \mathcal{D}^c \mathcal{D}^c$ for $m_{\mathcal{D}^c} > 500$ GeV (which means K-factor is greater than 2). 

\bibliography{refer}
\bibliographystyle{jhep}

\end{document}